\documentclass{article}
\usepackage{epsfig,subequation,graphics}
\catcode`\@=11
 
\newcommand{\Lvet}{\mbox{\boldmath $L$}}
\newcommand{\lvet}{\mbox{\boldmath $l$}}

\newcommand{\tvet}{\mbox{\boldmath $T$}}

\newcommand{\ov}{{\cal O}}

\newcommand  \f  \varphi
\newcommand \bra {\langle}
\newcommand \ket {\rangle}
\newcommand{\be}{\begin{equation}}
\newcommand{\ee}{\end{equation}}
\newcommand{\ben}{\begin{displaymath}}
\newcommand{\een}{\end{displaymath}}
\newcommand{\ba}{\begin{eqnarray}}
\newcommand{\ea}{\end{eqnarray}}
\newcommand{\ban}{\begin{eqnarray*}}
\newcommand{\ean}{\end{eqnarray*}}
\newcommand{\cro}{\dagger}

\newlength{\www}

\newcommand{\kt}{\mbox{\boldmath $k$}_{\perp }}

\newcommand{\Phivet}{\mbox{\boldmath $\Phi$}}

\def\lesssim{\mathrel{\mathop
  {\hbox{\lower0.5ex\hbox{$\sim$}\kern-0.8em\lower-0.7ex\hbox{$<$}}}}}
\def\greatsim{\mathrel{\mathop
  {\hbox{\lower0.5ex\hbox{$\sim$}\kern-0.8em\lower-0.7ex\hbox{$>$}}}}}
\relax

\begin{document}
\vspace{1cm}

\begin{titlepage}

\vspace*{-64pt}
\begin{flushright}
{  DFAQ-TH/2000-03\\
INFN-FE/99-???-P\\ 
hep-ph/0103316 \\}
\today
\end{flushright}

\vskip .7cm

\begin{center}
{\Large \bf 
Enhanced Electroweak Corrections to Inclusive Boson Fusion 
\\
Processes 
at the TeV Scale
\footnote{Work supported in part by EU QCDNET contract
FMRX-CT98-0194 and by MURST (Italy).}}
\vskip .7cm

{\large Marcello Ciafaloni}

{\it Dipartimento di Fisica, Universit\`a di Firenze,
\\ INFN - Sezione di Firenze, and CERN, Geneva\\
E-mail: ciafaloni@fi.infn.it}

\vskip.5cm

{\large Paolo Ciafaloni}

{\it INFN - Sezione di Lecce,
\\Via per Arnesano, I-73100 Lecce, Italy
\\ E-mail: Paolo.Ciafaloni@le.infn.it}

\vskip.5cm

{\large Denis Comelli}

{\it INFN - Sezione di Ferrara,
\\Via Paradiso 12, I-35131 Ferrara, Italy\\
E-mail: comelli@fe.infn.it}
\end{center}

\vskip .5cm

\begin{abstract}

Electroweak radiative corrections with double-log enhancements occur in 
inclusive observables at the TeV scale because of a lack of compensation
of virtual corrections with real emission due to the nonabelian (weak isospin) 
charges of the accelerator beams. Here we evaluate such Bloch-Nordsieck
violating corrections in the case of initial longitudinal bosons, which is
experimentally provided by boson fusion processes, and is related to the 
Goldstone-Higgs sector. 
All four states of this sector are involved in the group structure of the
corrections, and cause in particular a novel double log effect due to
hypercharge mixing in the longitudinal states. 
We study both the light- and the heavy-Higgs
cases, and we analyze the symmetry breaking pattern of the corrections.
The latter turn out to be pretty large, in the 5-10 \% range, and show an
interesting Higgs mass dependence, even for processes without Higgs boson in
the final state. 

\end{abstract}

\end{titlepage}

\setcounter{footnote}{0}
\setcounter{page}{0}
%

\section{Introduction}

Recent developments \cite{cc1}-\cite{ku} 
in the treatment of EW radiative corrections at 
high energies have brought attention to the fact \cite{3p,full}  
that double-log
 enhancements of infrared/collinear origin are present even in inclusive 
observables associated to hard processes at the TeV scale.
Such enhancements, involving the effective coupling
 $\alpha_W \log^2[s/M_W^2]$, signal the existence of a lack of compensation
 of the mass singularities associated with the $M^2_W/s\rightarrow 0$ limit,
 due to the non abelian weak-isospin charge of the initial states 
(that is, electrons or protons) provided by the accelerator.
In other words, the Bloch-Norsieck (B-N)theorem, valid for abelian theories, 
is here violated.

For instance, if we take  the example of $e e^+$ annihilation into hadron
jets in a Next Linear Collider (NLC\cite{NLC}), we found \cite{3p} that 
EW corrections to the total cross section 
at the TeV threshold are about $5 \%$  compared to the strong corrections 
of $3 \%$
and may even be larger if polarization and angular dependence are considered.
The simple fact which underlies this  surprising result is that
 the emission of a W boson by the incoming electron changes it into a neutrino 
whose hadron production cross section off
positrons is different from the electron-positron one, 
and actually happens to be larger than the latter.
This causes a lack of compensation with virtual corrections, 
and thus the  double-logarithmic enhancement which turns out
to be large in the TeV region.
Evaluation of such effects for the next generation colliders is thus needed.

It is amusing to note that the electroweak sector of the standard 
model is the first instance where such  non abelian effect is likely 
to be seen.
In fact, since QCD is in a confined phase, no colored asymptotic states
 are allowed: as a consequence,
colour averaging eliminates the double-log 
singularities and reduces the violation of collinear factorization to
 higher twist level \cite{dft}.
In the EW case, on the other hand, symmetry breaking allows the 
(non abelian) initial states to be prepared by electromagnetic forces, 
with ensuing violation of the B-N theorem and of collinear factorization
 theorems at leading level.
We remark elsewhere \cite{abelian} that a similar effect can occur in
 {\rm broken abelian theories} too, 
due to initial state mixing allowed by symmetry breaking. 

The enhancements of EW radiative corrections due to mass singularities
 were actually
 first pointed  out in exclusive processes at double \cite{cc1} and 
single \cite{lecce} logarithmic level.
In view of their size, resummations to all orders are
 available in the literature \cite{kuhn, cc2, fadin, ku} but
 we feel that no well-established treatment of higher order Sudakov
 logarithms has emerged
 yet in the broken symmetry case, especially for the subleading ones.
Here we stress again the point that the inclusive case is special 
 because it does not depend on 
the photon  effective mass, whose abelian effects cancel out by the 
B-N theorem.
Therefore, exponentiation of leading logarithms can be  established in 
a clearcut way \cite{3p}.

In a previous paper \cite{full} we have generalized the evaluation of
 the B-N violating logarithms to
 the case of initial gauge bosons, by limiting ourselves to  transverse
 polarizations.
Gauge bosons close to the mass shell are provided indirectly by the 
accelerator
in the form of boson-fusion processes, whose importance increases
 with the available energy
 in comparison with annihilation type processes.
We have thus provided a classification of the non canceling logs, 
which are roughly
 determined by the Sudakov form factor in the t-channel weak isospin
 representations occurring 
in the overlap matrix of the process.
The latter are found from the Clebsch Gordan couplings of the t-channel 
isospin to the weak
 isospin states of the initial particles, whether doublets 
(for fermionic initial states) 
or triplets (for bosonic ones).
This classification is here summarized in Sec.2.

In the present paper, we address 
 the problem of longitudinal initial bosons
 which is special
 because the longitudinal projection (Sec.3) singles out, by the equivalence
 theorem \cite{equiv}, 
the Higgs-Goldstone boson sector, which is sensitive to symmetry
 breaking and in particular
 to the Higgs boson mass $M_H$.

Since we work in the limit where the hard process invariants are in fixed  
ratios, and $s \gg M_W^2$, we are allowed in the fermionic and
 transverse boson cases to
 assume symmetry restoration for the squared matrix elements.
In the longitudinal case we need more care: if $M_H \simeq M_W$ 
(as hinted at by LEP data \cite{Lep}),
 then symmetry restoration is safely assumed for energies above
 the weak scale  (Sec.4).
On the other hand, if instead $M_H \gg M_W$, in the energy 
region $M_H \gg \omega \gg M_W$
the $SU(2)_L \otimes U(1)_Y$ symmetry is broken by the mass terms,
 and in general no easy 
treatment is allowed.
For this reason, we shall consider in detail the case of very small 
$\tan \theta_W=g'/g$,
 because in the $\theta_W \rightarrow 0$ limit the custodial $SU(2)_L
 \otimes SU(2)_R$
symmetry plays a key role \cite{cust}, and allows in fact a relatively simple 
treatment of the 
B-N violating double logs (Sec.5) in the energy region
$M_H \gg \omega \gg M_W$ also.

Finally, we discuss in Sec. 6 the example of jet production by boson fusion,
by providing both hard cross section and enhanced radiative corrections for 
the cases of a light or a heavy Higgs boson. A few details on the collinear
cutoffs and on the symmetry breaking pattern are derived in the Appendices.

\section{Electroweak double Logs in inclusive Processes}

EW radiative corrections with leading (double-log) enhancements
 occur in inclusive 
observables because of a lack of compensation of infrared logarithms between 
 virtual corrections and real emission.
This is due in turn to the fact that initial states in a (non abelian)
 multiplet
 may change flavour during emission, and thus interaction probability.
Such enhancements, therefore, have been classified \cite{full} 
according to the 
weak isospin 
properties of the initial states (fermions or bosons) and
 the resulting picture is summarized here.

We consider the structure of ``soft'' (infrared/collinear) interactions
 accompanying
 a hard standard model process $\{ \alpha_I\; p_I\}\rightarrow 
\{\alpha_F\; p_F \}$ 
where the $\alpha$'s and $p$'s denote collectively the flavours/colours 
and momenta
 of initial and final states.
The corresponding S-matrix can be written as an operator in the soft
 Hilbert space 
${\cal H}$, that collects the states which are almost degenerate with 
the hard
 ones, and as  a matrix in the hard labels, with form \cite{3p}
\be
 \label{eq:2}
S={\cal U}^F_{\alpha_F\beta_F}\! (a_s,a^\cro_s)\;\;\;
S^H_{\beta_F\beta_I}\! (p_F,p_I)\;\;\;
{\cal U}^I_{\beta_I\alpha_I}\! (a_s,a^\cro_s)
\ee
where ${\cal U}^F$ and ${\cal U}^I$ are operator functionals of the
soft emission operators $a_s,a^\cro_s$.

\begin{figure}
      \centering
      \includegraphics[height=80mm]
                  {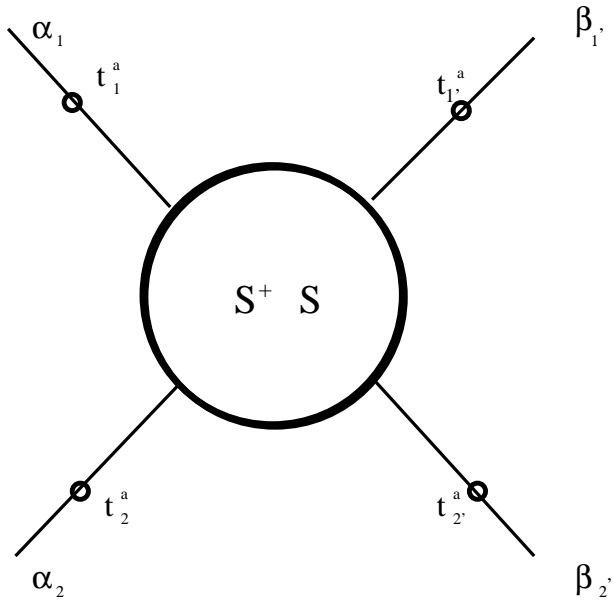}
      \caption{\label{overlap} Overlap Matrix}
      \end{figure}

An inclusive measurement on the hard process at hand involves
squaring the matrix elements in (\ref{eq:2})  and summing over the 
degenerate set $\Delta(p_F)$ of soft final states.
In this procedure, the final state interaction operators ${\cal U}^F$
 cancel out by unitarity, so
that the relevant quantity is the overlap matrix (Fig. \ref{overlap})
\be
\label{O}
\ov_{\beta_I \alpha_I}=\;_S \bra 0|
{\cal U}^{I\cro}_{\beta_I\beta'_I}(S^{H\cro}S^H)_{\beta'_I\alpha'_I}
{\cal U}^I_{\alpha'_I\alpha_I}|0 \ket_S
\ee
where $|0 \ket_S$ denotes the state without soft quanta, as is appropriate for 
the initial state.
It is also convenient to define the hard (tree-level) overlap matrix
 ${\cal O}^H=S^{H\cro}S^H$,
which  corresponds to eq. (\ref{O}) if soft 
radiative corrections are turned off.

Note that physical cross sections of initial particles which are
 eigenstates of flavours 
(weak isospin, charge, etc...) involve diagonal overlap matrix elements only.
However, off diagonal elements ($\beta_I\neq \alpha_I$) are also relevant 
in the case of  mixing (as for the $\gamma$ and $Z$ mass eigenstates) and in 
order to describe radiative 
corrections.
For this reason we consider a general label assignment in eq. (\ref{O}).

The factorization of soft operators assumed in eqs. (\ref{eq:2}) and (\ref{O}) 
is typical of 
a double log treatment, in which collinear singularities are considered in the
 soft limit, and  makes it clear that the B-N violating
double logs are due to the initial state interaction.
Generalization to subleading EW logarithms requires instead a critical
revision of the collinear
 factorization statements which are valid for QCD.

In order to explicitly evaluate the inclusive double logs, we have used
 \cite{3p} 
the method of asymptotic dynamics \cite{cm} which builds up the 
${\cal U}$-operators
 in (\ref{eq:2}) in terms of an effective large-time Hamiltonian,
 that is constructed in turn by an iterative method to all orders in the 
gauge couplings.
At leading double log level, and in the high energy limit $s \gg M_W^2
\simeq M_Z^2 \equiv M^2$,
the key ingredient of the asymptotic Hamiltonian is the eikonal soft-boson
 emission current \cite{cm}
\be \label{J}
J^{\mu a}(k)=g \;\sum_i \; \tvet^a_i\; \frac{p^{\mu}_i}{p_i \cdot k},
\;\;\;\;\;\;
J^{\mu 0}(k)=g' \;\sum_i \;y_i\; \frac{p^{\mu}_i}{p_i \cdot k}
\ee
which describes the emission of the gauge bosons eigenstates
$A^{\mu a},\;A^{\mu 0}$ ($a$=1,2,3) of
 momentum $k$ off the fast charges $p_i^{\mu}$ of energy 
$E_i\sim\sqrt{s}$ 
in the soft boson energy region $\sqrt{s} \gg \omega_k \gg M$.

Since we work in the high energy fixed angle limit for the hard process,  
we can assume that the $SU(2)_L \otimes U(1)_Y$ symmetry is
 restored in the squared matrix
 elements for $\sqrt{s} \gg M$.
Furthermore, since the inclusive double-logs induced by the e.m. 
charge $Q$ cancel out by the B-N theorem
for abelian theories, there is no net effect from the energy region
 $\omega_k<M$
 and we can work with only one cutoff
\be
\Theta_W=\theta(2 \,p\cdot k-M^2)
\ee
and with unmixed indices ($a$,0) in eq. (\ref{J}).

The isospin $ \tvet^a$ and hypercharge $y$ operators in eq. (\ref{J}) 
act on the flavour indices $\alpha_I\equiv(\alpha_1\;\alpha_2)$ of the
 (two) initial particles, which are doubled, because of the square,
 in the overlap matrix $\ov_{\beta_1\beta_2, \alpha_1 \alpha_2}$.
Because of the complex conjugation in the square, we shall normally use
 the representations
 $\tvet\;(-\tvet^T)$ for doublets on the $\alpha$-indices
 and antidoublets on the $\beta$-indices
 (antidoublets on the $\alpha$-indices and doublets on the $\beta$-indices),
 with the identifications 
\be
\label{os}
\bra \beta_1 \beta_2| \ov^{(s)}|\alpha_1\alpha_2 \ket=
\bra \bar{ \alpha}_2\beta_2| \ov^{(t)}|\alpha_1 \bar{\beta}_1 \ket=
\bra \ov|  \bar{\beta}_1 \bar{\beta}_2 \alpha_1\alpha_2 \ket
\ee
where the bars denote the conjugate representation
and the overlap matrix in the first  and the second  case
 is decomposed
 along the $s$  and $t$ channel flavour structure.

Restoration of gauge symmetry means that the gauge charges are conserved, i.e.,
\be\label{cons}
\ov (\sum_i \tvet^a_i)=\ov (\sum_i y_i)=0, \;\;\;\;(s\gg M^2)
\ee
and, as a consequence, the eikonal current (\ref{J}) is conserved too
\be
\ov \; J^{\mu A}(k) \;k_{\mu}=0\;\;\;\;\;\;\;\;(A=(a,0))
\ee
Residual effects of symmetry breaking come, of course, from the weak 
scale cutoff $M$, and from the fact that initial particles may be mixed, 
being a superposition of different isospin or hypercharge eigenstates.
By taking into account soft emission off the legs $\alpha_1,\; \beta_1$ 
(with momentum $p_1$) and 
$\alpha_2,\; \beta_2$ (with momentum $p_2$) the  eikonal current for the 
overlap matrix becomes,
by the gauge charge conservation (\ref{cons}):
\be
J^{\mu a}(k)=g (\frac{p_1^{\mu}}{p_1 \cdot k}-\frac{p_2^{\mu}}{p_2 \cdot k})
\tvet_L^a,\;\;\;\;\;\;\;\;
J^{\mu 0}(k)=g' (\frac{p_1^{\mu}}{p_1 \cdot k}-\frac{p_2^{\mu}}{p_2 \cdot k})
Y
\ee
where $\tvet_L^a=\tvet^a_1+\tvet^a_{1'}$ ($Y=y_1+y_{1'}$) denotes the total
  t-channel isospin (hypercharge).For initial states  with definite 
hypercharge ($y_1=-y_{1'}$  and $y_2=-y_{2'}$) we can set $Y=0$.
(But see sec.4 for a relevant example with $Y\neq 0$).

Radiative corrections to the hard overlap matrix (Fig.2) come from virtual
 contributions 
(connecting two $\alpha$-indices or two $\beta$-indices) and from
real emission (connecting an  $\alpha$-index  with a $\beta$-index).
A simple counting shows that a soft-boson loop contributes the eikonal
 radiation factor
\be
\frac{1}{2}\;(J^{ a}(k))^2=-g^2 \frac{p_1\cdot p_2}{(p_1\cdot k)(p_2\cdot k)}
\; \tvet_L^2\; \Theta_W
\ee
integrated over the phase space of the (on shell) soft boson.

\begin{figure}
      \centering
      \includegraphics[height=80mm]
                  {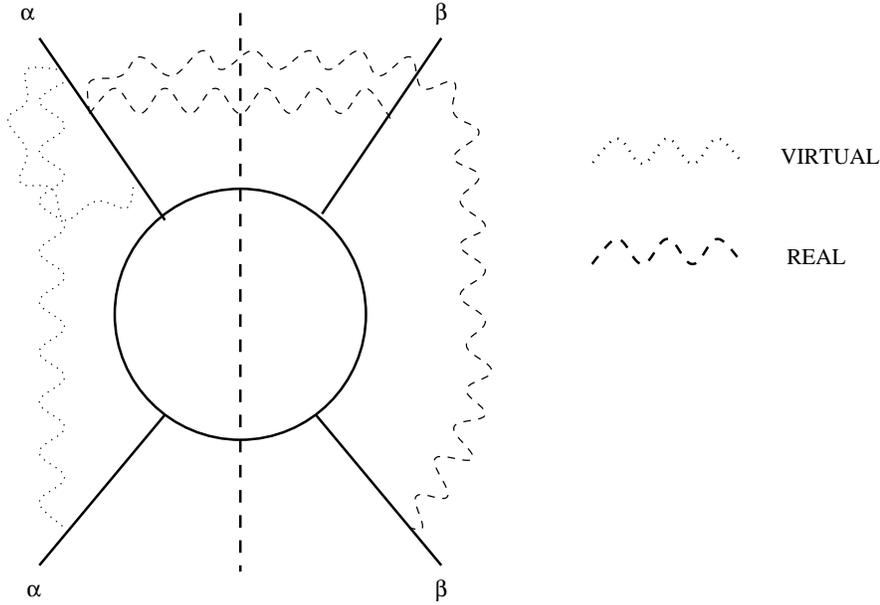}
      \caption{ One loop  Dressing of the Overlap Matrix}
      \end{figure}

The leading asymptotic hamiltonian is then found
by defining an evolution variable which is conjugated to the asymptotic time 
and by slicing the $k$ phase space accordingly \cite{cm}:
\be \label{Htau}
{\cal H}(\tau)=
\frac{g^2}{(2 \pi)^3} \int \frac{d^3 k}{2\, \omega} \;\delta(\tau-\tau(k))\;
 \frac{p_1\cdot p_2}{(p_1\cdot k)(p_2\cdot k)}\; \tvet_L^2\; 
\Theta_W \equiv
\tvet_L^2 \;\dot{\Lvet}_W(\tau)
\ee
In general, depending on the choice of $\tau$ whether energy \cite{cm} 
or emission
 angle \cite{angle} or $k_T$ variable \cite{kT},
higher order terms in the effective Hamiltonian may  be not negligible.
Here we stress the point that the present problem is commutative, ${\cal H}
(\tau)$ being proportional to
$\tvet_L^2$
at all  values of $\tau$, and thus any definition of $\tau$ will provide the
 same result, which is path ordering independent.
By $\tau$ integration we find in fact the Sudakov factor 
\be\label{sud}
S(s,M_W^2)=P\;e^{-\int d \tau {\cal H}(\tau)}=e^{-\Lvet(s,M^2)\,
\tvet_L^2}
\ee
where we have set
$\Lvet_W\equiv \Lvet (s,M^2)\equiv \frac{\alpha_W}{4 \pi}
\log^2\frac{s}{M^2}$,  which 
depends on the casimir $ \tvet_L^2$ of the total t-channel weak isospin of
 the problem.  Because of gauge symmetry, the hard overlap matrix 
${\cal O}_H$ has the form of a
sum of projectors of definite (t-channel) 
$SU(2)\otimes U(1)$ quantum numbers. It is then
straightforward to find the resummed expression for ${\cal O}$:
\be\label{new}
{\cal O}_H=\sum_{t_L} C_{t_L}P_{t_L}\qquad
{\cal O}=e^{-\Lvet(s,M^2)\tvet_L^2}{\cal O}_H=
\sum_{t_L} C_{t_L}e^{-\Lvet(s,M^2)t_L(t_L+1)}P_{t_L}
\ee

The inclusive double logs are thus found for a generic
initial state by the following steps:
\begin{itemize}
\item First, choose an inclusive hard process which is flavour blind (for 
instance $e^-e^+\rightarrow$ hadronic jets, or flavoured jets, summed over
 flavour charges).
\item Then, calculate the hard overlap matrix and its couplings to t-channel
 isospin.
\item Finally, get the radiative corrections by the weights (\ref{sud})
on each isospin state.
\end{itemize}

We found \cite{full}  by this procedure the following classification.

\subsection{Fermionic Initial States}

This is the most interesting case, which includes electron, 
muon, and proton colliders above the TeV scale.

Since left-handed fermions have $t_L=1/2$, we should
 consider doublet scattering for both the $ff$ 
(which is equal to $\bar{f}\bar{f}$) and $f\bar{f}$
(equal to $\bar{f}{f}$) cases.
Isospin and CP conservation provides the constraints 
\be\label{shivaree}
\sigma_{11}=\sigma_{22}=\sigma_{\bar{1}\bar{1}}=\sigma_{\bar{2}\bar{2}}
\;\;\;\;\;\;\;
\sigma_{1\bar{1}}=\sigma_{2\bar{2}}
\;\;\;\;\;\;\;
\sigma_{1\bar{2}}=\sigma_{\bar{1}2}
\ee
so that there are only four independent cross sections, e.g.,
$\sigma_{12},\;\;\sigma_{22}$ in the $ff$ case, and 
$\sigma_{1\bar{2}},\;\;\sigma_{2\bar{2}}$ in the $f\bar{f}$ one.

On the other hand, the t-channel states $|\alpha_1\bar{\beta}_1 \ket$
occurring in eq. (\ref{os})
 give rise to $\tvet_L=0,1$ for both the $ff$ and $f\bar{f}$ cases,
 thus providing again four independent coefficients \cite{full}:
\be
\ov_{ff}=C_0 \; P_0+C_1\;P_1\;\;\;\;\;\;\;
\ov_{f\bar{f}}=\bar C_0 \; \bar P_0+\bar C_1\; \bar P_1
\ee
Expressions for the resummed cross sections can be directly obtained by
projecting eq. (\ref{new}) on the subset of the states appearing in the
projectors that, having $t^3_L=0$, correspond to physical cross sections
\footnote{The elements of the overlap matrix corrisponding to physical cross
section always involve a given particle on leg 2  and its own antiparticle on
leg 2' in fig. \ref{overlap}, 
and have thus all t-channel additive quantum numbers set to 0.}. 
There
are two such states in the case at hand, providing the equations:
\be\label{new2}
\bra ii^*|{\cal O}|t_L=0,t_L^3=0\ket=\bra ii^*|{\cal O}^H|t_L=0,t_L^3=0\ket
\qquad
\bra ii^*|{\cal O}|t_L=1,t_L^3=0\ket=\bra ii^*|{\cal O}^H|t_L=1,t_L^3=0\ket
e^{-2\Lvet_W}
\ee
where $i=1,2,\bar{1},\bar{2}$ label the initial states on leg 1 and 1' 
in Fig. \ref{overlap}. Namely, we find:
\begin{itemize}
\item
The singlet $\tvet_L=0$ state
\be
\frac{1}{\sqrt{2}} (-|1 \bar 2\ket+|2 \bar 1  \ket )=
\frac{1} {\sqrt{2}} (|1 1^*  \ket +|2 2^* \ket)
\ee
corresponding to the average cross sections
\be
\frac{1}{{2}} (\sigma_{12}+\sigma_{22}),\;\;\;\;\;\;\;
\frac{1}{{2}} (\sigma_{1\bar 2}+\sigma_{2\bar 2})
\ee

\item
The triplet $\tvet_L=1$ states whose $\tvet_L^3=0$ component is 
\be
-\frac{1}{\sqrt{2}} (|1 \bar 2  \ket + |2 \bar 1 \ket)=
\frac{1} {\sqrt{2}} (|1 1^*  \ket -|2 2^* \ket)
\ee
corresponding to the cross-section differences
\be
\sigma_{12}-\sigma_{22},\;\;\;\;\;\;\;
\sigma_{1\bar 2}-\sigma_{2\bar 2}
\ee
\end{itemize}

By eq. (\ref{new2}), the average cross section has no radiative
 corrections, while the
 difference is suppressed by the form factor in the adjoint
 representation, e.g.,
\be\label{ee}
\sigma_{\nu e^+}-\sigma_{e^-e^+}=(\sigma^H_{\nu e^+}-
\sigma^H_{e^-e^+})\;\;e^{-2\Lvet_W}
\ee
This leads to corrections $\Delta \sigma_{e^-e^+}/\sigma^H_{e^-e^+}$ 
of order $5 \%$ for 
$e^-e^+$ annihilation into hadrons at the TeV treshold.

\subsection{Transverse Bosons}

This case is needed in order to treat boson-fusion processes, 
in which nearly on-shell gauge bosons are exchanged.
It must be supplemented by the longitudinal case to the studied 
below, which is favoured at very high energies.

Since transverse bosons behave as genuine isospin triplets, it 
is straightforward to
derive the isospin conservation constraints:
\be
\sigma_{++}=\sigma_{--},\;\;\;\;\;
\sigma_{3+}=\sigma_{3-},\;\;\;\;\;
\sigma_{33}=\sigma_{++}+\sigma_{-+}- \sigma_{3+}\;\;\;\;\;
\ee
so that only three independent cross sections are left
\be
\sigma_{\alpha}=\sigma_{\alpha+}\;\;\;\;\;\;\;\; (\alpha=+,3,-)
\ee
Correspondingly, the t-channel isospin can take the values 
$\tvet_L=0,1,2$, leading to 
three projectors  and three independent coefficients.

The $|\alpha_1\bar{\beta}_1 \ket$ states diagonalizing such projectors are:
\begin {itemize}
\item
The singlet state
\be
\frac{1}{\sqrt{2}}(|+- \ket+|-+ \ket+|33\ket)
\ee
corresponding to the cross section average 
$\frac{1}{3} \sum_{\alpha}\sigma_{\alpha}$;
\item
Three $\tvet_L=1$ states, whose  $\tvet_L^3=0$ component is
\be
\frac{1}{\sqrt{2}}(|+- \ket - |-+ \ket )
\ee
corresponding to the difference $\sigma_+-\sigma_{-}$;
\item
Five $\tvet_L=2$ states, whose  $\tvet_L^3=0$ component is
\be
\frac{1}{\sqrt{6}}(|+- \ket + |-+ \ket-2\,|33\ket)
\ee
corresponding to the cross section  combination 
\be\label{com}
\sigma_++\sigma_{-}-2\,\sigma_3
\ee
\end{itemize}

Due to the weights (\ref{sud}), the average cross section has
 no radiative corrections, the difference
 $\sigma_+-\sigma_{-}$ is damped by the form factor in the 
adjoint representation 
$\sim e^{-2 \Lvet_W}$,
much as in eq. (\ref{ee}), and the combination
(\ref{com}) is damped even more strongly, by the form factor in the
 $\tvet_L=2$ representation
 $\sim e^{-6 \Lvet_W}$.

\section{Soft Emission in the longitudinal Sector and Equivalence Theorem}

\begin{figure}
      \centering
      \includegraphics[height=60mm]
                  {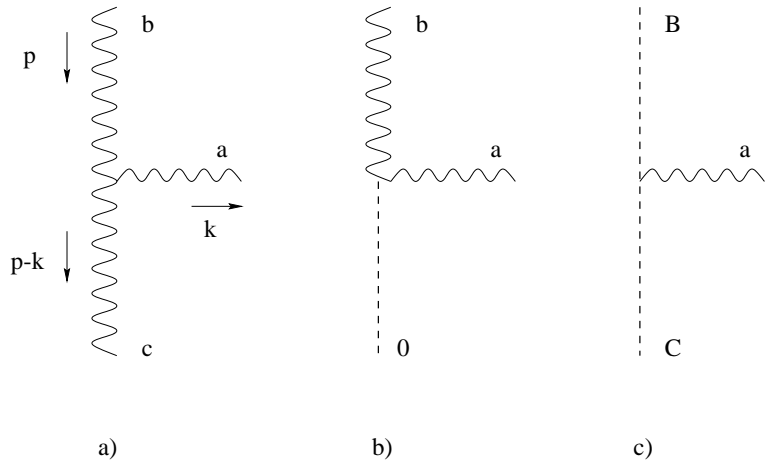}
      \caption{ Leading Eikonal Vertices in the Longitudinal 
(a,b) and Scalar (c) Sector }
      \end{figure}

It is well known that, in the high energy limit longitudinal 
polarizations are of type $\epsilon_L^{\mu}\simeq p^{\mu}/M+O(M/E)$,
and therefore longitudinal boson amplitudes $A_{\lambda}^a$ 
are related to the Goldstone bosons ones by the equivalence 
theorem \cite{equiv}
\be \label{equiv}
\frac{p^{\lambda}}{M}\;{\cal M}_{\lambda}^a(p ;...) \simeq
 \; i {\cal M}(\f^a(p);...)
,\;\;\;\;\;\;\;\;\;p^2\simeq M^2
\ee
where $a=1,2,3$ labels the $SU(2)_L$ vector boson indices and 
the remaining momentum and flavour indices of the amplitudes have
 been dropped.
\footnote{
Strictly speaking, we should have written eq. (\ref{equiv})
for the Z-boson instead of its $A^3$ component
$\frac{p^{\lambda}}{M_Z}\;{\cal M}_{\lambda}^{Z_L}(p ;...) \simeq
 \; {\cal M}(\f^3(p);...)$.
But using the relation $M_Z=M_W/c_W$ and electromagnetic current
 conservation provides eq. (\ref{equiv}) close to the vector boson mass shell
}
Therefore, the soft emission properties of the longitudinal bosons
 are related to the ones of Goldstone bosons, which are coupled to
 the Higgs boson in the scalar sector.
This shows that the dynamics of soft boson emission involves actually
 four real states $(\f^a,\;h)$ or the Higgs doublet and its conjugate.
We shall see that depending on the mass parameters of the problem,
 the doublet/antidoublet
or triplet/scalar description may be best suited, both  in general
 being needed.

In order to better understand this point, let us consider
  the emission of a soft boson with isospin index $a$, 
Lorentz indices $\mu$, and momentum $k$ off
 an energetic vector boson ($\epsilon_{\lambda}(p))$ via the 
three boson coupling.
The amplitude (Fig.3$a$) is proportional to
\ba
\epsilon_{\lambda}(p) \Gamma^{\lambda \mu \nu}(p,k)&=&
\epsilon_{\lambda}(p)
[g^{\lambda \nu}\;(2\,p+k)^{\mu}-g^{\mu \nu} \;(p+2\, k)^{\lambda}\\
&-&g^{\lambda 
\mu}\;(p-k)^{\nu}]
\simeq 2\, p^{\mu}\,\epsilon^{\nu}(p)-\epsilon^{\mu}(p)\, p^{\nu}+O(k)
\ea
where use has been made of the condition $\epsilon(p) \cdot p=0$.
By taking into account the denominator and isospin factors, we then
 obtain a generalized eikonal emission amplitude
\be 
\bra {\cal M}(b,\epsilon(p);...)\;|a,\mu,k\ket
\simeq i\, g\, \epsilon_{bac}\;
\left[{\cal M}(c,\epsilon(p+k);...) 
\frac{p^{\mu}}{p \cdot k}-
\frac{M}{2\, p \cdot k}\epsilon^{\mu}(p)\;
{\cal M}(c,\epsilon_L(p+k);...)\;\right]
\ee
which differs from the customary one of the unbroken theories by 
the second term in square brackets, which is purely longitudinal.
If $\epsilon(p)$  is a transverse polarization, the second
 term is of order $\frac{M}{E}$ with respect to the first one,
 and is thus subleading.
On the other hand, if $\epsilon(p)\sim p/M+O(M/E)$
is the longitudinal polarization,
 then the latter term cancels one half the ``leading'' term, and leads to
a factorized eikonal emission current off longitudinal bosons:
\be \label{ampl}
\bra {\cal M}(b,\epsilon_L(p);...)\;|a,\mu,k\ket
\simeq i\, \frac{g}{2}\, \epsilon_{bac}\;
\left [{\cal M}(c,\epsilon_L(p+k);...)\;\frac{p^{\mu}}{p \cdot k}
\;(1+0(\frac{\omega}{E}))\right]
\ee
Note the factor of $\frac{1}{2}$, relative to the normal eikonal 
factorization theorem for
transverse  polarizations.

The amplitude (\ref{ampl}) is not however, the only possible
 contribution to soft boson emission, which can occur via a 
Higgs boson also (Fig.3$b$).
Taking into account the weak scale dependence properly, one finds
 again the factorized eikonal current, with a Higgs contribution 
on the r.h.s.:
\be \label{am}
\bra {\cal M}(b,\epsilon_L(p);...)\;|a,\mu,k\ket
\simeq \frac{g}{2} \,\frac{p^{\mu}}{p \cdot k}\;
\left[
i\, \epsilon_{bac}\;
  {\cal M}(c,\epsilon_L(p+k);...) -
i\, \delta_{ab}\;
{\cal M}(h(p+k);...)\;
\right]
\ee
Since the Higgs boson can occur as a final state as well (Fig.3$c$)
 it is convenient to generalize eq. (\ref{am}) in matrix form on the
 boson indices $B=(b,0)\equiv
(\f_a,h)$, as follows:
\be \label{amp}
\bra {\cal M}(\f^B(p);...)\;|a,\mu,k\ket
\simeq  g\,\frac{p^{\mu}}{p \cdot k}\;
(\tvet_L^a)_{BC}\;{\cal M}(\f^C(p+k);...) 
\ee
by thus introducing the weak isospin matrix in the $(\f^a,\,h)$ basis
\be\label{T}
\tvet_L^a=\frac{1}{2}\;\left(\begin{array}{cc}
i\, \epsilon_{bac}& -i\, \delta_{ab}\\
 i\, \delta_{ac}&0\end{array}\right)
\equiv\frac{1}{2}(T^a_V+T^a_h)
\ee
where
$(T^a_V)_{bc}=i\, \epsilon_{bac}$ .
Similarly one can define the eikonal emission current 
of the $B_{\mu}\equiv A_{\mu}^0$ boson in the form:
\be
\bra {\cal M}(\f^B(p);...)\;|0,\mu,k\ket
\simeq  g'\,\frac{p^{\mu}}{p \cdot k}\;
Y_{BC}\;{\cal M}(\f^C(p+k);...) 
\ee
where
\be\label{Y}
Y=\frac{1}{2}(T^3_V-T^3_h)
\ee
is the hypercharge matrix in this sector.
Eqs. (\ref{T}) and (\ref{Y}) are the basis for our treatment of
 radiative corrections in the following.

It is  important to realize that the gauge generators 
(\ref{T}) and (\ref{Y}) have  a simple structure within the custodial
 symmetry group $
SU(2)_L \otimes SU(2)_R$ \cite{cust} .
In fact it is straightforward
to check that $T_V$ and $T_h$ in eq. (\ref{T}) satisfy the algebra of
 the $O(4)$ group
\be
[T_V^a,T_V^b]=i\, \epsilon^{abc}\, T_V^c,\;\;\;\;
[T_V^a,T_h^b]=i\, \epsilon^{abc}\, T_h^c,\;\;\;\;
[T_h^a,T_h^b]=i\, \epsilon^{abc}\, T_V^c,\;\;\;\;
\ee
As a consequence, the $SU(2)_L$ ($SU(2)_R$)
generators $\tvet_L^a$ ($\tvet_R^a$) are provided by
\be
\tvet_L^a=\frac{1}{2}(T^a_V+T_h^a),\;\;\;\;\;\;\;
\tvet_R^a=\frac{1}{2}(T^a_V-T_h^a)
\ee
So that $[\tvet_L^a,\tvet_R^b]=0$ and $Y=\tvet_R^3$.
In this language the basis $(\f^a,h)$ is just the $(\frac{1}{2},\frac{1}{2})$ 
representation of the group
$O(4)\sim SU(2)_L \otimes SU(2)_R$, characterized by $\tvet_L^2=\tvet_R^2=3/4$,
 and the matrices (\ref{T}) and (\ref{Y}) refer to the 4-vector representation.

Of course, the same type of eikonal current  (\ref{amp}) arises by direct
 study of the Goldstone-Higgs sector and of their couplings to soft 
bosons, as expected from the equivalence theorem (Fig.3$c$).
The longitudinal projection just provided helps clarifying some aspects 
of the problem, in  particular the factor $\frac{1}{2}$ occurring in (\ref{T}),
 and the occurrence of four states associated with the longitudinal bosons.
It is sometimes useful to arrange such  four states of the 
$(\frac{1}{2},\frac{1}{2})$
 representation in the complex form of the doublet/antidoublet matrix
\be\label{high}
\Phivet=
\frac{1}{\sqrt{2}}(h+i\tau_a\f_a)
=\left(\begin{array}{cc}
\f_0&i\f_+\\i\f_-&\f_0^*\end{array}\right)
=\frac{1}{\sqrt{2}}\left(\begin{array}{cc}
h+i\f_3&\f_2+i\f_1\\-\f_2+i\f_1&h-i\f_3\end{array}\right)
=\left(\begin{array}{cc}
\bar{1}&{1}\\\bar{2}&{2}\end{array}\right)
\ee
which transforms as
$
\Phivet\rightarrow U_L\Phivet U_R^\cro
$
under the $ SU(2)_L \otimes SU(2)_R$
group. The indices $1,2,\bar{1},\bar{2}$ here refer to the transformation
properties under SU(2)$_L$.

\section{Radiative Corrections in the light Higgs case}

For Higgs bosons and vector bosons  nearly degenerate, the gauge 
symmetry of the squared matrix elements is restored in the full
 high-energy region $\omega \gg M= M_W\simeq M_H$ (the common weak scale),
 while the photon scale plays no role because of the B-N theorem for QED.
Therefore, we can assume $SU(2)_L \otimes U(1)_Y$
symmetry for both the hard overlap matrix and for
enhanced radiative  corrections, as already discussed in sec.2.

Since the longitudinal bosons are related to the Higgs doublets by the
 equivalence theorem (sec.3), and the latter have $t_L=1/2$, it would 
seem that we are in the same case as that of fermion and antifermion
doublets.
However, a peculiarity of the longitudinal/Higgs  sector is the fact that 
the physical Higgs boson is by itself a manifestation of (low-energy)
 symmetry breaking, being a coherent superposition (see (\ref{high}))
\be\label{higgs}
|h \ket=\frac{|\f_0\ket+|\f_0^*\ket}{\sqrt{2}}=
\frac{|\bar{1}\ket+|2\ket}{\sqrt{2}}
\ee
of doublet states of opposite hypercharge.
Since the state (\ref{higgs})
 has no definite hypercharge, it happens
 that overlap matrix elements having non vanishing t-channel
 hypercharge $Y=\pm 1$ occur
in physical cross sections initiated by neutral states, besides the 
normal ones with $Y=0$.

The fact above has no counterpart  in the light fermion case 
\footnote{Note  however that mixing occurs in the case 
of transversely polarized fermions, which 
  are indeed a superposition of states of opposite chirality, and of
  different gauge quantum numbers \cite{abelian}}
because it
 is not possible to prepare a coherent superposition of a fermion and
 of its antifermion, due to exact fermion number conservation.
On the other hand, a similar phenomenon occurs in the customary case 
of $B_{\mu}-W_{\mu}^3$ mixing yielding $\gamma$ and $Z$ mass eigenstates, as
already discussed in ref.\cite{full} .
In the latter case the mixing is nonabelian, involving $t_L=1$ and $t_L=0$
 states, while in the Higgs case the mixing is abelian, involving
 $y=\pm1/2$ hypercharge states.
A peculiar consequence of this fact is that violation of B-N theorem 
can occur in broken abelian theories too \cite{abelian}.

By using the definitions (\ref{high}), 
the gauge symmetry constraints (\ref{shivaree})
in the longitudinal /Higgs sector can be written as follows:
%
\be\label{++}
\sigma_{--}=\sigma_{++},\;\;\;\;\;;\;
\sigma_{+3}=\sigma_{-3}=\sigma_{+h}=\sigma_{-h}
\ee
In the purely neutral sector we find, on the other hand, the mixing
 phenomenon, in the form:
\ba\label{00}
\sigma_{33}&=&\sigma_{hh}=\frac{1}{2}(\sigma_{++}+\sigma_{+-})+{\rm Re} 
(\ov)\\
\nonumber
\sigma_{3h}&=&\frac{1}{2}(\sigma_{++}+\sigma_{+-})-{\rm Re} (\ov)
\ea
where $\ov\equiv \ov(\f_0\f_0^* \rightarrow \f_0^* \f_0)$  and 
its conjugate denote the $Y=1$ and $Y=-1$ overlap matrix elements
 mentioned before.

It is then convenient to refer to a fixed target (the $+$ state, say)
in order to enumerate the independent cross sections.
Due to eqs. (\ref{++})
we find at once three of them 
\be \label{+3}
\sigma_{++},\;\sigma_{-+},\;
\sigma_{3+}
\ee
 similarly to the transverse polarization case  and in addition to
 the neutral overlap $Re\ov$ just defined.
Finally, we also need another off diagonal matrix element 
\be\label{h3}
I\equiv {\rm Im} \ov_{h3}\equiv 
{\rm Im}(\ov_{3+\rightarrow h+})=\frac{1}{2}(\sigma_{0 +}-
\sigma_{0^* +})
\ee
in order to express all possible initial state dependences.
All these five quantities, except $Re\ov$, have $Y=0$ in the t-channel.

We shall now compute the enhanced (B-N violating) radiative 
corrections to the observables 
(\ref{00}) and (\ref{+3}).
Following our method (sec.2), we have to 
(i) define
the hard process and the inclusive (flavour blind) 
observable we are interested
in 
(ii) derive the lowest order (tree level) hard overlap matrix
 $\ov^H_{\beta_1\beta_2,\alpha_1\alpha_2}$ and 
(iii) derive
the effective hamiltonian describing the enhanced radiative corrections,
 for the given initial state configuration.

Symmetry considerations restrict the hard overlap matrix, according to
 the possible t-channel quantum numbers whose eigenstates are constructed
 out of the 
$|\alpha_1\bar{\beta}_1\ket$ states defined before.
In the longitudinal/Higgs sector $\alpha_1$ and $\beta_1$ span the four states
of the $(1/2,1/2)$ representation, so that the 
$|\alpha_1\bar{\beta}_1\ket$  states
can be classified as follows (see Appendix B for the explicit expressions):
\begin{itemize}
\item Eight ($Y=0)$ states, which can be grouped according to left
 isospin representation, and $CP=\pm1$ eigenvalues.
We find thus the projectors:
\be\label{state1}
T^{CP}(m)=0^{-}(1),\;\;1^-(3),\;\;   0^+(1)    ;\; \; 1^+(3),
\ee
corresponding  to four independent coefficients, each with the
 multiplicity $m$ of states in parenthesis.
Such four projectors are in close correspondence with the ones for 
the fermionic doublets discussed in sec 2.
Note that the $CP$ eigenstates $+1$ ($-1$) are just symmetric 
(antisymmetric) under doublet-antidoublet interchange.
\item Four ($Y=1)$ 
and four ($Y=-1)$
states, which are interchanged by $CP$ and must thus be degenerate,
 which can be classified according to left isospin representation:
\be\label{state2}
T(m)=0(2),\;\;   1(6) 
\ee
corresponding to two more independent coefficients.
\end{itemize}

Actually only five out of the overall six independent coefficients are coupled 
to the observables enumerated in eq. (\ref{+3})  and (\ref{h3}),
 (the one corresponding to $\tvet_L=0$ and
$Y=Q = \pm 1$ being decoupled from physical states by the e.m.
 charge superselection rule).

The effective hamiltonian describing radiative corrections is found
 by exactly the same method as in sec 2, with the important difference
 that now we can have a non vanishing t-channel hypercharge,  and we
 thus obtain
\be\label{HH}
{\cal H}(\tau)=\dot{\Lvet}_W(\tau)(\tvet_L^2+tg^2 \theta_W\;Y^2)
\ee
from which the t-channel Sudakov factors
\be
S(\tvet_L,Y)=e^{-\Lvet_W(s,M_W^2)(\tvet_L^2+tg^2\theta_W\;Y^2)}
\ee
are found for each one of the states (\ref{state1}) and (\ref{state2}).

In order to assign the Sudakov weights to the observables (\ref{++})
 and (\ref{h3})
 we need to relate them to the explicit form of the states which diagonalize
 the projectors  (\ref{state1}) and (\ref{state2}).
A direct construction of them is obtained from their classification 
according the $ SU(2)_L \otimes SU(2)_R$
 group; this construction is explicitly worked out in Appendix B, 
From the direct product of two $(\frac{1}{2},\frac{1}{2})$
representations we obtain the following representations
\be\label{LR}
(\frac{1}{2},\frac{1}{2})\otimes(\frac{1}{2},\frac{1}{2})
=(0,0)\oplus (0,1)\oplus(1,0)\oplus(1,1)
\ee
which  are related to those in (\ref{state1})
and (\ref{state2}) as follows:
\begin{itemize}
\item 
The state $(0,0)$ is the $0^-$ state and corresponds to
 the average cross section
\be \label{id}
\sigma_++\sigma_{-}+\sigma_0+\sigma_{\bar{0}}=
\sigma_++\sigma_{-}+\sigma_3+\sigma_{h}=
\sigma_++\sigma_{-}+2\,\sigma_3\equiv \Sigma+2\,\sigma_3
\ee
where we introduce the notation $\sigma_{\alpha}\equiv 
\sigma_{\alpha +}$ in which the $+$  target is understood.
The identification (\ref{id}) is obtained by the explicit construction 
\ba\label{a}
|0;0\ket&=&\frac{1}{2}(-|1 \bar{2}\ket+
|2 \bar{1}\ket+| \bar{1} 2 \ket -
| \bar{2} 1\ket)=\frac{1}{2}(|+-\ket+
|-+\ket+
| 0 \bar{0}  \ket +
| \bar{0} 0\ket)=\\
&&
\frac{1}{2}(|+-\ket+
|-+\ket+
| 33   \ket +
| hh \ket)
\ea
which has, by inspection, $t_L=t_R=0$, and is also antisymmetric
 into the doublet indices interchange $i\leftrightarrow \bar{i}$.
Eq.(\ref{id}) corresponds to the matrix element
$\bra +-|\ov|0;0\ket $.
\item 
The three states $(0,1)$ are split according to the hypercharge 
$Y=\tvet_R^3$
into (i) the two degenerate $t_L=0,\;Y=\pm 1$ states in (\ref{state2}) and 
(ii) the symmetric $t_L=Y=0$
state in (\ref{state1}).
The latter corresponds to the combination
\be
\sigma_+ - \sigma_{-}+
2\; {\rm Im} \ov_{h3}\equiv \Delta+2\;I,\;\;\;\;\;\;
 \ov_{h3}=\bra h+|\ov|3+\ket,\;\;\;\;\;\;
0^+
\ee
and to the state $0^+$, provided by
\be\label{b}
|0;t_R=1,Y=0\ket=\frac{1}{2}(-|1 \bar{2}\ket+
|2 \bar{1}\ket - | \bar{1} 2 \ket +
| \bar{2} 1\ket)=\frac{1}{2}
(|+-\ket -
|-+\ket+
| 0 \bar{0}  \ket -
| \bar{0} 0\ket)
\ee
\item
The three states $(1,0)$ are the antisymmetric $1^-$ states, whose 
$t_L^3=0$ term is the only
one related to cross-sections, and corresponds to the combination
\be
\sigma_+ - \sigma_{-}-2\; Im \ov_{h3}\equiv \Delta-2\;I,\;\;\;\;\;\;1^-
\ee
and to the state
\be\label{c}
|1, t^3_L=0;0\ket=\frac{1}{2}(-|1 \bar{2}\ket-
|2 \bar{1}\ket + | \bar{1} 2 \ket +
| \bar{2} 1\ket)=\frac{1}{2}
(|+-\ket -
|-+\ket-
| 0 \bar{0}  \ket +
| \bar{0} 0\ket)
\ee

\item
The nine $(1,1)$ states are split, according to the hypercharge
 quantum number into (i) six degenerate $t_L=1,Y=\pm 1$ states
occurring in (\ref{state2}) and (ii) three $Y=0,\;\;1^+$ states
 occurring in (\ref{state1}), whose $t^3_L=0$ 
term corresponds to the cross section
\be
\sigma_++\sigma_{-}-\sigma_0-\sigma_{\bar{0}}=
\sigma_++\sigma_{-}-\sigma_3 -\sigma_{h}=
\sigma_++\sigma_{-}-2\,\sigma_3\equiv \Sigma-2\,\sigma_3,\;\;\;\;\;\;
(1^+)
\ee
and to the state
\be
|1,t^3_L=0;1,t^3_R=0\ket=\frac{1}{2}
(-|1 \bar{2}\ket-
|2 \bar{1}\ket-
| \bar{1} 2 \ket -
| \bar{2} 1\ket)=
\frac{1}{2}(|+-\ket+
|-+\ket-
| 0 \bar{0}  \ket -
| \bar{0} 0\ket)
\ee
On the other hand, the state corresponding to the overlap
 $\ov(\f_0\f^*_0\rightarrow \f^*_0\f_0)$
in the neutral sector has $t^3_L+t^3_R=Q=0$ and corresponds to
the states $|00\ket$ or $|\bar 0\bar 0\ket$ in (\ref{state1}).
Other states with $Y\neq0$ give rise to $Q\neq0$ and therefore 
cannot be coupled to physical cross sections.
\end{itemize}

The cross-section assignments 
just provided are summarized in Table 1(a,b).

We are now in a position to provide the Sudakov evolution in eq. (\ref{HH})
for each one of the cross section combinations just introduced.
By using the relevant $(t_L,Y)$ assignments, we find
\ba\label{system1}
\sigma_+ &+&\sigma_{-}+2\,\sigma_3=\Sigma^H+2\,\sigma^H_3,
\;\;\;\;\;\;\;\;\sigma_h=\sigma_3\\ \nonumber
\sigma_+ &+&\sigma_{-}-2\,\sigma_3=(\Sigma^H-2\,\sigma^H_3)\;e^{-2 \Lvet_W},\\
\nonumber
\sigma_+ &-&\sigma_{-}+2\,I=\Delta^H+2\,I^H,\\\nonumber
\sigma_+ &-&\sigma_{-}-2\,I=(\Delta^H-2\,I^H)\;e^{-2 \Lvet_W},\\\nonumber
{\rm Re}\ov&=&{\rm Re}\ov^H\;e^{- \Lvet_W(2+ tg^2 \theta_W)},\\\nonumber
\sigma_{3h}&=&\frac{1}{2}(
\sigma_+ +\sigma_{-})-{\rm Re} \ov,\\\nonumber
\sigma_{33}&=&\sigma_{hh}=\frac{1}{2}(
\sigma_+ +\sigma_{-})+{\rm Re} \ov
\ea
from which all radiative corrections can be derived once the hard 
(tree level) cross sections are known.

\section{The heavy Higgs Case and Symmetry Breaking Pattern}

If the Higgs mass happens to be much larger than the vector boson mass,
 then in the energy region $M_H\gg \omega\gg M_W$ gauge symmetry is broken
 by the mass spectrum even if the overall energy of the process is $s\gg
 M_H^2$.
    Therefore, if we want to keep track of the leading powers of all
     logarithms, whether $\log[M^2_H/M_W^2]$ or the customary ones
     $\log[s/M_W^2]$, we end up with a process with two mass scales, in which
     symmetry breaking causes a lack of conservation of the gauge current and
     therefore requires a more refined treatment of the B-N violation logs and
     of their scales.

     Since we feel that despite recent efforts \cite{ku} the
     direct diagrammatic calculation of higher order Sudakov logarithms in the
     many scale, broken symmetry case, is not yet fully understood, we prefer
     to treat here the particular case $g'/g\rightarrow 0$, in which the
     custodial symmetry $SU(2)_L\otimes SU(2)_R$ of the longitudinal/Higgs
     sector plays a key role, by providing a conserved current which
     considerably simplifies the problem.  Note in fact that, at tree level
     and in the high energy limit $s \gg M_H^2 \gg M_W^2$, the gauge symmetry
     is still valid and is actually extended to the full $SU(2)_L\otimes
     SU(2)_R$ group in the $\theta_W \rightarrow 0$ limit.  On the other hand,
     radiative corrections in the energy region  $M_H \gg \omega \gg M_W$
     break the gauge symmetry, but keep the global custodial symmetry $SU(2)_V
     \subset  SU(2)_L\otimes SU(2)_R$ which then provides the conserved
     current we are interested in.

     In order to understand this point, recall
     that the infrared/collinear  behavior is regulated by the boson emission
     current (\ref{amp})  where the weak isospin charge is written as the sum
     of its vector  component ($1/2\; T_V^a$) and of a Higgs contribution.  In
     order to take into account the mass cutoffs in the denominators, we write
     the current in eq. (\ref{amp}) in the more precise form \be \label{currh}
     J^{a \mu}(k)=g \frac{p^{\mu}}{p \cdot k}\frac{1}{2}(T^a_V\; \Theta_W+
     T^a_h\; \Theta_H)=g \frac{p^{\mu}}{p \cdot k}(\tvet^a_L\;
     \Theta_H+\frac{1}{2} T^a_V\; \Theta_W(1-\Theta_H)) \ee where
 \be\label{sl} \Theta_i=\theta(2 p\cdot k -M^2_i),\;\;\;\;\;\;\;i=W,H \ee
 denote the mass cutoffs.  

  It is apparent from eq. (\ref{currh}) that the soft
 boson insertion  current  is the full gauge current for $2 p\cdot k>M^2_H$,
 while it becomes the $SU(2)_V$ current (with coupling $g/2$) in the region
 $M_H^2>2 p\cdot k>M^2_W$.  Since each current is conserved in the
 corresponding validity region,  the evolution of radiative corrections
 follows the same pattern as that outlined in sec.2, and the evolution
 hamiltonian derived from the current  (\ref{currh}) becomes 
\be \label{hamh}
 {\cal H}(\tau) =\dot{\Lvet}_H(\tau)\;\tvet^2_L+\frac{T_V^2}{4}
 \,(\dot{\Lvet}_H(\tau)-\dot{\Lvet}_W(\tau)) \ee where the
 $\dot{\Lvet}_H(\tau),\;\;(\dot{\Lvet}_H(\tau)-\dot{\Lvet}_W(\tau))$
 contribution is defined with the phase space slicing in eq. (\ref{Htau}),
 following the cutoff structure $\Theta_H,\;\;(\Theta_H-\Theta_W)$ in
 eq.(\ref{currh}). After $\tau$-integration, for $E>M_H^2/M_W$, the two 
 exponents turn out to be
 (Appendix A): \be \Lvet_W=\frac{\alpha_W}{4 \pi}\; \log^2\frac{s}{M^2_W}
 \;\;\;\;\;\;\; \Lvet_H=\frac{\alpha_W}{4 \pi}\;( \log^2\frac{s}{M^2_W}-2\,
 \log^2\frac{M^2_H}{M^2_W})\equiv \Lvet_W-2\lvet_h \ee 

 Note that, since $T_V^a=\tvet^a_L+\tvet^a_R$, $\tvet^a_L$
 transforms as a vector  under $T_V$ rotations, and therefore $\tvet^2_L$
 transforms as a scalar , i.e., it is  invariant.  It follows that the two
 terms in  (\ref{hamh}) commute between each other, and therefore $[{\cal
 H}(\tau),{\cal H}(\tau')]=0$ thus  making the Sudakov exponent independent of
 the type of phase space slicing (whether in $\omega$ or in $k_T$).  By $\tau$
 integration we then obtain \be\label{sudh}
 S(s,M_H^2,M_W^2)=e^{-\tvet^2_L\,\Lvet_H-\frac{T_V^2}{4}\,
 (\Lvet_W-\Lvet_H)} \ee Note finally that the hamiltonian
 (\ref{hamh}) commutes with $T_V^a$ (but not with $\tvet^a_L$), so that the
 overlap matrix radiative corrections are diagonalized by $SU(2)_V$, as
 anticipated  before.

  In practice, explicit projectors and cross section
 evolution are constructed as follows.  The hard overlap matrix posseses 
full  $
 SU(2)_L\otimes SU(2)_R$ symmetry, and therefore is classified directly
 according to the representation occurring in the direct product (\ref{LR}),
 corresponding to four independent coefficients.  If radiative corrections are
 turned on, the symmetry is broken to $SU(2)_V$ and the corresponding states
 are classified according to the Casimir $T_V^2$.  Eigenstates of $T_V^2$ are
 trivially constructed for the cases
\begin{itemize}
\item[a)] $\tvet^2_L=T_V^2=0,\;\;\; CP=-1$
\item[b)] $\tvet^2_L=0,\;\;T_V^2=\tvet_R^2=1,\;\;\; CP=+1$
\item[c)] $\tvet^2_L=T_V^2=1,\;\;\; CP=-1$

discussed before, because they are directly provided by 
eqs.(\ref{a}),(\ref{b}),(\ref{c}), respectively.

In the case 

\item[d)] $\tvet_L=\tvet_R=1$

the degeneracy is resolved by the vector sum $T_V^a=\tvet_L^a+\tvet_R^a$,
 so that we can have $T_V=0,1,2$.
By straightforward algebra (appendix B) we find
\begin{itemize}
\item[d0)] The $\tvet_L=1,\;\;T_V=0$ state

\ba \label{55}
\frac{1}{\sqrt{3}}(
-|\bar 1\bar1 \ket 
-|22\ket 
-\frac{1}{2} (
|1 \bar 2\ket +
|2 \bar 1\ket +
|\bar   1 2  \ket +|\bar 2 1 \ket))=\\\nonumber
\frac{1}{2\sqrt{3}}
(-2\, |00 \ket -2\, |\bar 0 \bar 0 \ket +
|+- \ket +
|-+ \ket - |0 \bar 0 \ket  - | \bar 0 0 \ket)
\ea
corresponding to the cross section combination
\be
\sigma_{+}+\sigma_{-} +\sigma_3-3\; \sigma_h
\ee

\item[d1)] Three $ \tvet_L=T_V=1$  states whose $Q=T_V^3=0$ term is 

$\frac{1}{i \sqrt{2}}(|\bar 1 \bar 1\ket -| 2 2\ket )=
\frac{1}{i \sqrt{2}}(|\bar 0 \bar 0 \ket -| 0 0 \ket ) $
corresponding to an overlap of the type
$Im \ov(\f_0 \f^*_0\rightarrow \f^*_0 \f_0)$ which 
does not occur in physical cross sections.

\item[d2)] Five  $ \tvet_L=1,\;\;T_V=2$
states, whose 
$T_V^3=0$ term is
\be\nonumber
\frac{1}{\sqrt{6}}(
|\bar 1\bar1 \ket 
+|22\ket 
-
|1 \bar 2\ket -
|2 \bar 1\ket -
|\bar   1 2  \ket -|\bar 2 1 \ket)=
\frac{1}{\sqrt{6}}
( |00 \ket + |\bar 0 \bar 0 \ket +
|+- \ket +
|-+ \ket - |0 \bar 0 \ket  - | \bar 0 0 \ket)
\ee
corresponding to the cross section combination
\be
\sigma_{+}+\sigma_{-} -2\;\sigma_3
\ee
\end{itemize}
The cross-section assignments just provided are summarized in Table 1(a,c).
\end{itemize}

Correspondingly, the independent cross sections are $\sigma_{+},\;
\sigma_{-},\; \sigma_3,\;\sigma_h$, while the neutral sector 
cross sections can be expressed in terms of the four preceding ones
 by the relationships
\ba\label{rel}
\sigma_{h3}&=&\sigma_{h+}=\sigma_{h}\\ \nonumber
\sigma_{33}&=&\sigma_{+}+\sigma_{-}-\sigma_3\\\nonumber
\sigma_{hh}&=&\sigma_{+}+\sigma_{-}+\sigma_3-2 \, \sigma_h
\ea
The relations on the first line of eq.(\ref{rel})
are the simple $SU(2)_V$ constraints already used in sec.2.
The last one is obtained by straightforward algebra by using the eigenstate 
(\ref{55})
on the $\tvet_L=1,\;\;T_V=0$ projectors (Appendix B).

The B-N violating corrections are finally found by assigning 
the Sudakov form factors  (\ref{sudh})  to each of the cross section
 combinations found before.
We obtain
\ba \label{system2}
\sigma_+ &+&\sigma_{-}+\,\sigma_3+\,\sigma_h=\Sigma^H+2\,\sigma^H_3,\\ 
\nonumber
\sigma_+ &+&\sigma_{-}+\,\sigma_3-3\,\sigma_h=(\Sigma^H-2\,\sigma^H_3)
\;e^{-2 \Lvet_H},\\\nonumber
\sigma_+ &-&\sigma_{-}+2\,I=(\Delta^H+2\,I^H) \;e^{-\frac{1}{2} (
\Lvet_W-\Lvet_H)} ,\\\nonumber
\sigma_+ &-&\sigma_{-}-2\,I=(\Delta^H-2\,I^H)\;e^{-2 \Lvet_W +\frac{3}{2} (
\Lvet_W-\Lvet_H)},\\ \nonumber
\sigma_+ &+&\sigma_{-}-2\,\sigma_{3}=(\Sigma^H-2\,\sigma_3^H)\;
e^{-2 \Lvet_W +\frac{1}{2} (
\Lvet_W-\Lvet_H)}
\ea
from which each cross section can be derived.

\begin{table}
{\Large
\centering
\begin{tabular}{||c||c||c||c||c||c||}\hline
 & ($t_L,\;t_R$) & $t_L^{CP}$ & $Y^2$ & $T_V$ & Cross Section\\\hline
a &(0,0)& $0^-$ & 0 &0& $\sigma_+ +\sigma_{-} +\sigma_3 +\sigma_h $\\\hline
a&(0,1)& $0^+$ & 0 &1& $\sigma_+ -\sigma_{-} +2\, Im\ov_{h3}$ \\\hline
b, c&(1,0)& $1^-$ & 0 &1& $\sigma_+ -\sigma_{-} -2\, Im\ov_{h3} $\\\hline
b &(1,1)& $1^+$ & 0 && $\sigma_+ +\sigma_{-} -\sigma_3 -\sigma_h$ \\\hline
c &(1,1)&  &  & 2 & $\sigma_+ +\sigma_{-} -2\, \sigma_3 $ \\\hline
b &(1,1)& $1$ & 1 && $ 2 \, Re \ov $ \\\hline
c &(1,1)&  &  & 0 & $\sigma_+ +\sigma_{-} + \sigma_3 -3\,\sigma_h  $ \\\hline
\hline
\end{tabular}
\par} 
\vskip 0.5 cm
\caption{t-Channel quantum number assignments of various 
cross section combinations.
(a) Invariants, (b) $SU(2)_L \otimes U(1)_Y$  states ($M_H \simeq M_W$) and
(c)  $SU(2)_V$ states ($M_H\gg  M_W$). 
The unphysical $Q\neq 0$ states have been omitted.
}\label{tab-cod:1}
\end{table}


\section{Phenomenological Applications}

Before applying the radiative corrections just found to some explicit 
phenomenological cases 
it is interesting to solve the systems (\ref{system1}) and
(\ref{system2}), and to investigate their asymptotic limit 
$ \Lvet_W\rightarrow \infty$. In such a limit 
the values of the evolved cross sections - with radiative corrections -
will depend only on some invariant combinations that do not evolve, 
i.e., are free of double-logs.

For the case with $\sin \theta_W\neq 0$ and $M_W\simeq M_H$
we have the two ``invariants''
\be
\frac{\sigma_++\sigma_{-}+\sigma_h+\sigma_3}{4}\equiv \Theta_0;\;\;\;\;
\frac{\sigma_+- \sigma_{-}+2\,I}{4}\equiv \Theta_1
\ee
with the constraint $\sigma_h=\sigma_3$. The first one corresponds to the 
cross-section average over the four Goldstone/Higgs states, and the second is
the difference between the doublet and antidoublet averages. As a consequence,
all cross sections converge to a combination of
such two invariants:
\ba
\sigma_{\pm}&\rightarrow&(\Theta_0^H \pm \Theta_1^H)\\\nonumber
\sigma_3,\; \sigma_{33}&\rightarrow&\;\Theta_0^H\\\nonumber
I &\rightarrow&\; \Theta_1^H\\\nonumber
{\rm Re} \ov &\rightarrow&0
\ea
where the superscript $H$ means that the invariants are evaluated at Born 
level.

For the case with $\sin \theta_W=0$ and $M_W\ll M_H$
we also have the same two ``invariants'', but now we find an extra 
$M_H$ dependence in the limit 
$ \Lvet_W\rightarrow \infty$ \footnote{
We stress that the limit that we are taking is
 $\sqrt{s}\rightarrow \infty$ with $M_H, \;M_W$ fixed.
In such a case the combination
$\Lvet_H -\Lvet_W=-2 \frac{\alpha_W}{4 \pi}\log^2[M_H^2/M_W^2]\equiv -2\; 
\lvet_h$ is finite.
}:

\ba
\sigma_{\pm}&\rightarrow&(\Theta_0^H \pm
 e^{-\lvet_h}\; \Theta_1^H)\\
\nonumber
\sigma_3,\; \sigma_{33} &\rightarrow&\;\Theta_0^H\\\nonumber
I &\rightarrow&\;  e^{-\lvet_h}\;\Theta_1^H
\ea
Therefore, in the extra limit $M_H \gg M_W$  all the physical cross sections
would converge to the invariant $\; \Theta_0^H$, i.e., the
average over all states.

The evaluation of the ratios $\sigma_{ij}/\Theta^H_i$ 
give very easily an idea of the 
strength of the evolution of the cross section
$\sigma_{ij}$  under investigation.
For instance, the cross section $\sigma_{33}$ is of interest, because it can
be identified in an $e^+e^-$ collider by a double tagged experiment where the
two final electrons both have transverse momenta $|\kt|^2\greatsim M^2$.
By eq. (\ref{system1}) it has the expression
\be\label{label}
\sigma_{33}=\frac{1}{4}[\Sigma^H+2\sigma_{3+}^H+
e^{-2\Lvet_W}(\Sigma^H-2\sigma_{3+}^H)+2{\rm Re}
{\cal O}^He^{-\Lvet_W(2+\tan^2\theta_W)}]
\ee
in the case $M_H\approx M$, and an analogous one in the 
case $M_H\gg M$. The limit $\sigma_{33}\to \Theta_0^H$ is obvious at large
energies, while the full behavior depends on the explicit values of the Born
cross sections.

\begin{figure}
      \centering
      \includegraphics[height=80mm]
                  {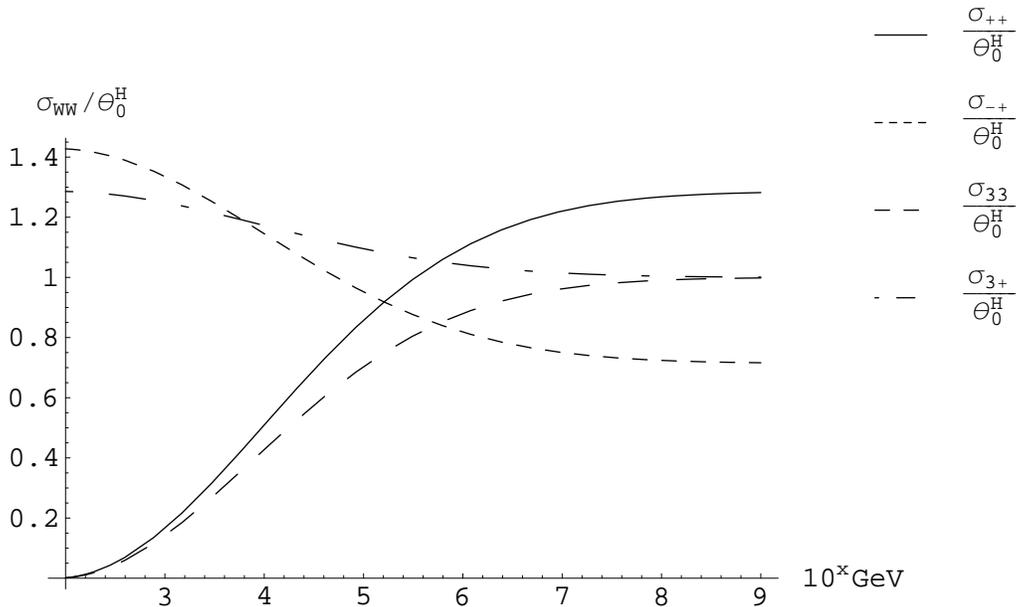}
      \caption{$W_LW_L\rightarrow  f \bar{f}$ inclusive cross section for
$M_W \simeq M_H$ }
      \end{figure}

Now let us discuss the example of $W_L W_L\rightarrow$ hadronic jets.
This cross section is relevant to describe hadron jet production at
 NLC by boson fusion, a process which becomes competitive with the 
annihilation process at high energies.

The Born cross sections for $W_L W_L\rightarrow q\bar{q}$ are well
 known \cite{renard}:

\ba
\frac{d \sigma^H_{++}}{d \cos \theta}&=&
\frac{d \sigma^H_{33}}{d \cos \theta}=0\\  \nonumber
\frac{d \sigma^H_{3+}}{d \cos \theta}&=&
\frac{d I^H}{d \cos \theta}=
\frac{\pi \alpha_W^2 N_c N_f}{8\, s}\;
\frac{\sin^2 \theta}{8}\\ \nonumber
\frac{d \sigma^H_{-+}}{d \cos \theta}&=&\frac{\pi \alpha_W^2 N_c N_f}{8\, s}\;
\frac{\sin^2 \theta\;(9-18 \,\sin^2 \theta_W+20\,\sin^4 \theta_W) }
{72\, \cos^4 \theta_W }
\ea
The corresponding evolution, e.g., for $\sigma_{33}$ in the case 
$M_H\approx M$ becomes, by eq. (\ref{label})
\be
\sigma_{33}=\frac{1}{4}[
\sigma_{-+}^H+2\sigma_{3+}+e^{-2\Lvet_W}(\sigma_{-+}^H-2\sigma_{3+}^H)-
2\sigma_{-+}^He^{-\Lvet_W(2+\tan^2\theta_W)}]\approx
\frac{1}{4}[\sigma_{-+}^H+2\sigma_{3+}+\tan^2\theta_W\sigma_{-+}^H]2\Lvet_W
\ee

In this case $\sigma_{33}$ becomes nonvanishing for $s>M^2$ because of
radiative corrections, and $\frac{\delta\sigma_{33}}{\Theta_0^H}\approx
2\Lvet_W$ is roughly of the order 10 \% at the TeV threshold. This provides an
estimate of the double logs, which are therefore of the same order as for
fermions  and transverse bosons initial states.

In  Fig.4 we plot the ratios $ \sigma_{ij}/\Theta^H_0$ 
($\sin \theta_W\neq 0,\;M_H\simeq M_W,\; \cos \theta=0$)
 over
a huge energy range to show the ``convergence speed''
 of the evolved cross sections.

In Fig.5 to the left 
we plot the same  ratios as before but with three different 
 $M_H=100,\;500,\;1000$ GeV cases
   ($\sin \theta_W= 0,\;M_H> M_W,\; \cos \theta=0$).
To the right we plot the ratio between a cross section calculated with 
$M_H=100$ GeV over the same cross section calculated  with 
$M_H=1000$ GeV to show the Higgs mass dependence as a function of the energy.
This figure  appears interesting from the Higgs physics point of view:
while the hard $WW\rightarrow \bar{f} f$
 cross section is independent of the Higgs mass at high energies,
 the evolved ones instead show a  pronounced
 effect coming from the exponential $e^{-\lvet_H}$
 which tends to depress the  corrections.
This means that also completely inclusive cross sections not involving
explicit Higgs production show enhanced Higgs mass dependence in their
radiative corrections.
Obviously, before  bringing this simple example into 
the realm of  phenomenology it needs  many other
inputs like luminosities of the longitudinal bosons, backgrounds  etc.  
In any case we think that this extreme example can be 
 useful to show the  potential
importance  of this kind of  ``new Standard Model effects''
related to  the measure of inclusive cross sections at very high energy.


\begin{figure}[htb]\setlength{\unitlength}{1cm}
\begin{picture}(12,7.5)
\put(0.3,1){\epsfig{file=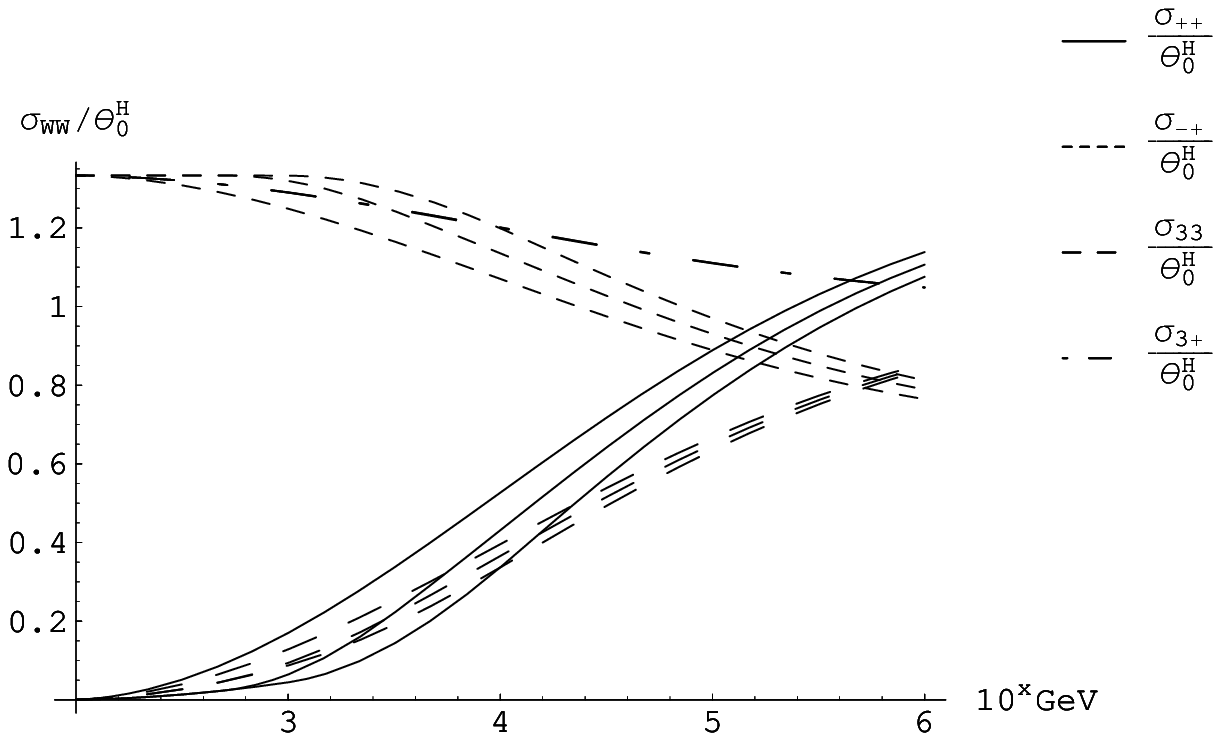,height=5cm}}
\put(9.3,1){\epsfig{file=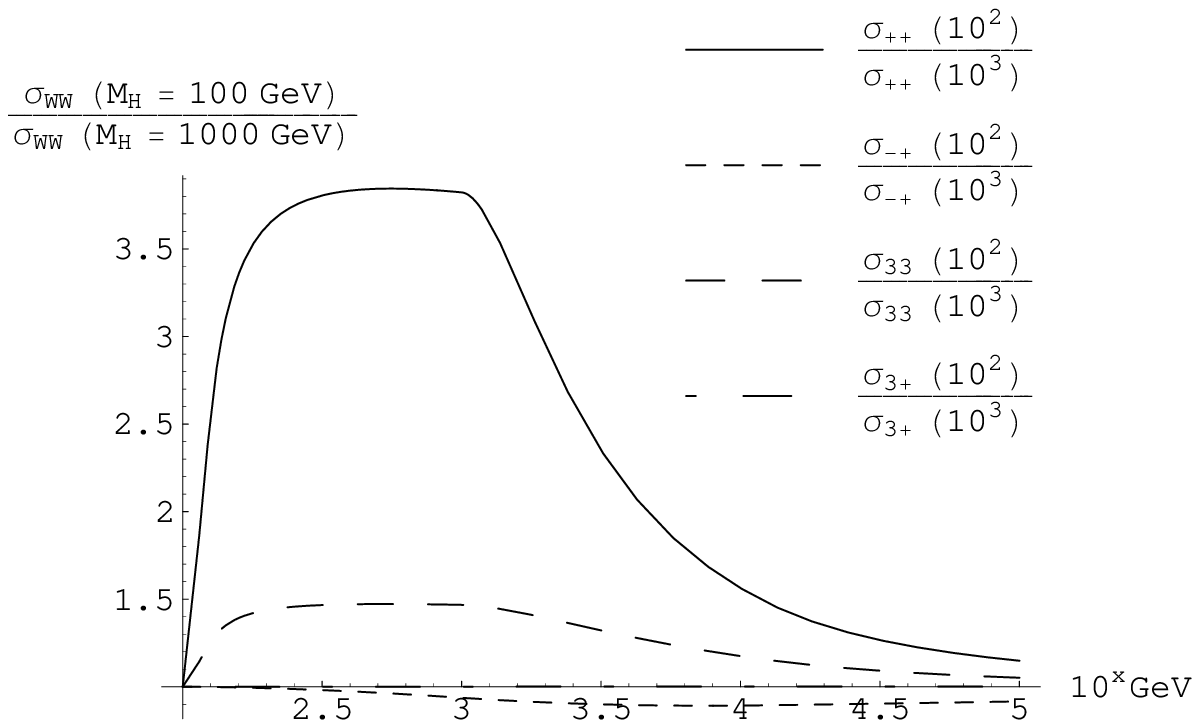,height=5cm}}
\end{picture}
\caption{$W_LW_L\rightarrow f \bar{f}$ inclusive cross sections for
$M_W < M_H$ and $\sin \theta_W=0$.
In the first picture to the left we show for 
three different values of $M_H=100,\;500,\;1000$ GeV
 the energy evolution of the cross sections.
On the right we plot the energy dependence of
 the ratio between the cross sections calculated with $M_H=100$ GeV
 over the ones calculated with $M_H=1000$ GeV.
}
\end{figure}

\section{Conclusions}
 
Our study of the longitudinal initial states occuring in boson fusion processes
has shown that, in this case, enhanced radiative corrections have a number of
interesting and sometimes surprising features. 

First of all, soft interactions involve all four states of the Higgs sector,
and this makes the group-theoretical classification of the corrections
somewhat more involved than in previously analyzed cases (Sec. 2). We have
provided here (Table 1) the combinations of cross sections with well defined
form factors, for both 
$M_H\approx M_W$ and $M_H\gg M_W$, and we have studied their radiative 
evolution.

Secondly, we find important {\sl abelian} double logs \cite{abelian}, due to 
the breaking of weak hypercharge, besides the nonabelian ones. This is
surprising at first sight, and is due to the occurence of off-diagonal
overlap matrix elements, corresponding to a nontrivial hypercharge in the
t-channel 
(eq. (\ref{system1})). In turn, this happens because
the longitudinal and 
Higgs states are mixed, i.e., are not hypercharge eigenstates.

Finally, the quantitative effects are sizeable, in the 5-10 \% range 
at the
TeV threshold as for the fermionic and transverse bosons initial states
investigated previously. Furthermore, because of the enhanced radiative
corrections, the Higgs mass dependence is not
negligible (Fig. 5) even for processes without Higgs bosons in the final
states, that are therefore insensitive to the Higgs mass at tree level.

On the whole, we have now provided a fairly complete description of the
enhanced EW corrections for fermionic and bosonic initial states, which can
now be used for detailed predictions of hard inclusive observables, 
at the
double log level, at future accelerators energies.
Of course, this is not enough: we actually need a generalization of collinear
factorization to broken gauge theories, in order to account for single
logarithms as well. 

\section{Appendix A: Soft Radiation Coefficients}

Here we derive the expansions of the evolution coefficients $\dot{\Lvet}_W,
\; \dot{\Lvet}_H$ and of their $\tau-$integrals for the explicit
 choice $\tau=\omega$ and $\tau=k_T$ of the evolution variable.

We start considering the effect of the collinear cutoff
\be \label{const}
\Theta_H=\Theta(2 pk-M_H^2)
\ee
on one of the incoming momenta (either $p=p_1$ or $p=p_2$), and the corresponding calculation of the evolution hamiltonian in eq.(\ref{Htau}).

By taking either $\omega$ or $k_T$ fixed the constraint 
(\ref{const}) is written as
\ba\label{A2}
&&\frac{1}{\omega}\;
Max\left(M,\sqrt{\frac{\omega}{M}}\,M_H\right)<\theta<1, \;\;\;\;\;\;\;
(\tau=\omega)\\
&& k_T<\omega<
Min\left(\frac{k_T^2 \sqrt{s}}{M_H^2},\; \sqrt{s}\right), \;\;\;\;\;\;\;
(\tau=k_T)
\ea
By then integrating the eikonal radiator factor in eq.(\ref{Htau}) over the region (\ref{A2}) we obtain
($E\equiv\sqrt{s}$)
\be
\frac{\pi\tau}{\alpha_W}\;\dot{\Lvet}_H=
2\,
\log\frac{\omega}{M}\;\Theta(E\,M^2-M_H^2\omega)+2
\, \log\frac{\omega\,E}{M_H^2}\;\Theta(\omega\,M_H^2-M^2\,E\omega)
\ee
for $M<\tau=\omega<E$, and
\be
\frac{\pi\tau}{\alpha_W}\;\dot{\Lvet}_H=
2\,
\log\frac{E}{k_T}\;\Theta(k_T-M_H)+2
\, \log\frac{k_T \,E}{M_H^2}\;\Theta(M_H-k_T)
\ee
for $M<\tau=k_T<E,\;\;\;\;E>M_H^2/M$, 
while $\dot{\Lvet}_W$ is given by the same expressions in the $M_H=M$ limit.

Finally, by $\tau-$integration, we obtain in both cases
\be
\Lvet_H=\frac{\alpha_W}{\pi}\left(log^2\frac{E}{M}-2\, 
log^2\frac{M_H}{M}\right), \;\;\;\;\;\;\;\;
E>\frac{M_H^2}{M}
\ee
which corresponds to eq.
(\ref{hamh}) in the text.
In the intermediate energy region $M_H<E<M_H^2/M$ we get instead
\be
\Lvet_H=2\,\frac{\alpha_W}{\pi}\;log^2\frac{E}{M_H}, \;\;\;\;\;\;\;\;
M_H<E< \frac{M_H^2}{M}
\ee
as expected for a smooth $E=M_H$ limit.


\section{Appendix B} 

The overlap matrix can always be written as a sum of projectors 
${\cal O}=\sum C_i P_i$. While the coefficients $C_i$ depend on the dynamics
and therefore on the physical process considered, the projectors $P_i$ can be
classified and explicitly constructed according to the relevant symmetry. 
We choose to classify the states according to their 
$SU(2)_L\otimes SU(2)_R$ symmetry properties.
This symmetry acts 
on the representation
$(\frac{1}{2},\frac{1}{2})$ as follows:
\be
\left(\bar{\Phi}\Phi\right)=
\left(\begin{array}{cc}
\bar{1}&1\\\bar{2}&2\end{array}\right)
\to\exp[i\alpha_L^a\tau^a]
\left(\Phi^c\Phi\right)
\exp[-i\alpha_R^b\tau^b]
\ee
so that
using the basis  $|t_L,t^3_L;t_R,t^3_R\ket$ we can do the 
following identification:
\be
|1\ket=|\frac{1}{2},\frac{1}{2};\frac{1}{2},\frac{1}{2}\ket\quad
|2\ket=|\frac{1}{2},-\frac{1}{2};\frac{1}{2},\frac{1}{2}\ket\quad
|\bar{1}\ket=|\frac{1}{2},\frac{1}{2};\frac{1}{2},-\frac{1}{2}\ket\quad
|\bar{2}\ket=|\frac{1}{2},-\frac{1}{2};\frac{1}{2},-\frac{1}{2}\ket
\ee

Then the states coming from the direct sum 
$T_L=t^1_{L}+t^{1'}_{L}$ and $T_R=t^1_{R}+t^{1'}_{R}$ 
generate the representations
$(\frac{1}{2},\frac{1}{2})\otimes(\frac{1}{2},\frac{1}{2})
=(0,0)+(1,0)+(0,1)+(1,1)$ that we label as 
$|T_L,T^3_L;T_R,T^3_R\ket$. Explicitly, the states with $(T_L,T_R)=(1,1)$
are given by:
\begin{equation}\label{1}\begin{array}{lll}
|1,-1;1,-1\ket=|\bar{2}\bar{2}\ket\qquad
&
|1,-1;1,0\ket=\frac{|2\bar{2}\ket+|\bar{2}2\ket}{\sqrt{2}}\qquad
&
|1,-1;1,1\ket=|22\ket
\\
|1,1;1,-1\ket=|\bar{1}\bar{1}\ket\qquad
&
|1,1;1,0\ket=\frac{|1\bar{1}\ket+|\bar{1}1\ket}{\sqrt{2}}\qquad
&
|1,1;1,1\ket=|11\ket
\\
|1,0;1,-1\ket=\frac{|\bar{1}\bar{2}\ket+|\bar{2}\bar{1}\ket}{\sqrt{2}}\qquad
&
|1,0;1,0\ket=\frac{|1\bar{2}\ket+|\bar{1}2\ket+
|2\bar{1}\ket+|\bar{2}1\ket}{2}\qquad
&
|1,0;1,1\ket=\frac{|{1}{2}\ket+|{2}{1}\ket}{\sqrt{2}}
\end{array}\end{equation}
The states $(T_L,T_R)=(1,0)$ are:
\begin{equation}\label{2}\begin{array}{lll}
|1,-1;0,0\ket=\frac{|2\bar{2}\ket-|\bar{2}2\ket}{\sqrt{2}}\qquad
&
|1,0;0,0\ket=\frac{|1\bar{2}\ket-|\bar{1}2\ket+
|2\bar{1}\ket-|\bar{2}1\ket}{2}\qquad
&
|1,1;0,0\ket=\frac{|1\bar{1}\ket-|\bar{1}1\ket}{\sqrt{2}}
\end{array}\end{equation}
while those $(T_L,T_R)=(0,1)$ are:
\begin{equation}\label{3}\begin{array}{lll}
|0,0;1,-1\ket=\frac{|1\bar{2}\ket-|\bar{2}1\ket}{\sqrt{2}}\qquad
&
|0,0;1,0\ket=\frac{|1\bar{2}\ket+|\bar{1}2\ket-
|2\bar{1}\ket-|\bar{2}1\ket}{2}\qquad
&
|0,0;1,1\ket=\frac{|12\ket-|21\ket}{\sqrt{2}}
\end{array}\end{equation}
Finally, the singlet $(T_L,T_R)=(0,0)$ is:
\be\label{4}
|0,0;0,0\ket=\frac{|1\bar{2}\ket-|\bar{1}2\ket-
|2\bar{1}\ket+|\bar{2}1\ket}{2}
\ee


In the $M_H\simeq M_W$ case, the gauge symmetry
 $SU(2)\otimes U(1)$ is the relevant one, and we can classify the states
according to the $(T_L,T^3_R=Y)$ possible values; however this does not
provide a complete classification of the states since $U(1)_Y$ is an abelian
symmetry. For instance,  we have two states 
corresponding to $T_L=0,Y=0$; one of them ($|0,0;0,0\ket$
in eq. (\ref{4})) 
is CP-odd while the other
 ($|0,0;1,0\ket$ in (\ref{3})) is CP-even. We can then use CP invariance to
further classify the states, obtaining:

Eight ($Y=0$) states, whose multiplicity $m$ is written in parenthesis:
\be\label{state1app}
|T_L,Y^2\ket^{CP}(m)=|0,0\ket^{-}(1),\;\; |0,0\ket^+(1) ,\;\;  
|1,0\ket^-(3),\;\;   |1,0\ket^+(3)
\ee

Four ($Y=1)$ 
and four ($Y=-1)$
states, which are interchanged by $CP$ and must thus be degenerate:
\be\label{state2app}
|T_L,Y^2\ket(m)=|0,1\ket(2),\;\;   |1,1\ket(6) 
\ee

We have then 6 independent projectors,  
classified by their $SU(2)\otimes U(1)$ and CP properties; 4 of them
correspond to $Y^2=0$ and are given in (\ref{state1app}) and the remaining 2
with
$Y^2=1$ are given in (\ref{state2app}). 
The subset of states corresponding to physical cross sections (with
$T^3_L=Y=0$) can be written in terms of the ones appearing in 
(\ref{1}-\ref{4}) as follows:
\be\begin{array}{ll}
|T_L=0,T^3_L=T^3_R=0\ket^+=|0,0;1,0\ket\quad
&
|T_L=0,T^3_L=T^3_R=0\ket^-=|0,0;0,0\ket\quad
\\
|T_L=1,T^3_L=T^3_R=0\ket^+=|1,0;1,0\ket\quad
&
|T_L=1,T^3_L=T^3_R=0\ket^-=|1,0;0,0\ket\quad
\end{array}\ee
Because of the mixing
phenomenon in the neutral sector (see text), we  also
need the subset of
(\ref{state2app}) with
$Q=T^3_L+T^3_R=0$ but $Y=T^3_R\neq 0$:
\be\begin{array}{ll}
|T_L=1,T^3_L=-1,T^3_R=1\ket=|1,1;1,-1\ket\quad
&
|T_L=1,T^3_L1,=T^3_R=-1\ket=|1,-1;1,1\ket\quad
\end{array}\ee

%




In the $M_H \gg M_W$ case
we need classify the states according to their $SU(2)_L\otimes SU(2)_R$
properties and their vectorial  isospin $\tvet_V=\tvet_L+\tvet_R$.
Writing them as $|T_L,T_R;T_V\ket$ we have then the possibilities:
\be
|0,0;0\ket\quad |0,1;1\ket\quad|1,0;1\ket\quad
|1,1;0\ket\quad |1,1;1\ket\quad|1,1;2\ket\quad
\ee
The first three of them are given directly by eqs. (\ref{4},\ref{3},\ref{2}),
the states with $T^3_L=T^3_R=0$ corresponding to physical cross sections. 
The remaining three can be found by using the usual rules of isospin
composition. It is relatively easy to find the states with
$Q=T^3_L+T^3_R=0$:
\be\begin{array}{ll}
|1,1;0\ket \supset
\frac{1}{\sqrt{3}} (|1,0;1,0 \ket + |1,1;1,-1\ket+
|1,-1;1,1\ket )
&
|1,1;1\ket \supset\frac{i}{\sqrt{2} } ( |1,1;1,-1\ket- |1,-1;1,1\ket )
\\
|1,1;2\ket \supset
\frac{1}{\sqrt{6}} (-2|1,0;1,0\ket +|1,1;1,-1\ket+
|1,-1;1,1\ket )&\end{array}
\ee

\end{document}